\documentclass{ifacconf}

\usepackage{graphicx}      
\usepackage{natbib}        
\usepackage{amsmath}
\usepackage{algorithm}
\usepackage{algpseudocode}
\usepackage{amssymb}
\usepackage{booktabs}
\usepackage{threeparttable}
\def\objective{J}

\usepackage{pgfplots}
\usepackage{multicol}
\usepackage{multirow}
\usepackage{eurosym}
\usepackage{subcaption}
\usepackage{microtype}
\usepgfplotslibrary{groupplots}
\usepackage{tcolorbox}

\ifodd 0
\newcommand{\rv}[1]{{\color{blue} #1}}
\else
\newcommand{\rv}[1]{{#1}}
\fi

\begin{document}
\begin{frontmatter}

\title{What price to pay? Auto-tuning a building MPC controller for optimal economic cost} 

\author[First]{Jiarui Yu} 
\author[First]{Jicheng Shi} 
\author[First]{Wenjie Xu}
\author[First]{Colin N. Jones}

\address[First]{Laboratoire d'Automatique, EPFL, 1015 Lausanne, Switzerland.}

\thanks{The first three authors contribute equally to this work. Corresponding author: Wenjie Xu, 
 e-mail address: wenjie.xu@epfl.ch.}

\begin{abstract}                
Demand-side management (DSM) programs introduce complex pricing, requiring advanced control for cost minimization. Model Predictive Control (MPC) offers a solution but its performance hinges on appropriate hyperparameter tuning. We propose using Constrained Bayesian Optimization (CONFIG) to automate this process. In a case study, our optimized MPC reduced electricity costs by 26.90\% compared to a rule-based controller and by 17.46\% versus an manually tuned MPC. Analysis of real contracts further showed that optimal DSM program selection can lower monthly bills by up to 20.18\%, demonstrating a data-driven path to significant consumer savings.
\end{abstract}


\end{frontmatter}

\section{Introduction}
The building sector is a major contributor to global energy consumption, accounting for approximately 34\% of total use~\citep{20.500.11822_41133}. In response to growing energy pressures, price-based demand-side management (DSM) programs~\citep{EID201615}, such as time-of-use and real-time pricing, have been widely adopted. These programs aim to enhance grid stability by incentivizing consumers to shift their energy use through fluctuating electricity prices.
While beneficial for the grid, DSM programs introduce complexity for consumers, who must navigate a wide array of electricity contracts and optimize their building's control strategy to minimize costs without sacrificing comfort. 
Existing contract selection services analyze historical usage patterns~\citep{su15118750} to recommend suitable contracts, but using historical data alone is insufficient. Greater saving requires adaptive control strategies tailored to specific pricing structures~\citep{WANG20241196}, which traditional rule-based controllers struggle to handle under dynamic pricing~\citep{YANG2020115147}.
Model Predictive Control (MPC) has emerged as a leading solution for this challenge due to its predictive capabilities. However, conventional MPC design process, which lacks hyperparameter tuning stage, often fails to deliver optimal performance in real-world conditions. Model inaccuracies and environmental disturbances can lead to a misalignment between prediction and control objectives, resulting in suboptimal performance~\citep{Mishra_2024}.

Several studies have explored performance-oriented controller design for buildings. For instance, \cite{XU2024122493} used a primal-dual contextual Bayesian optimization method to tune a PI controller, reducing energy use while maintaining comfort. Similarly, \cite{buildings13020314} optimized HVAC setpoints via Bayesian optimization, and \cite{REN2024122312} employed iterative optimization for multi-energy storage deployment. However, these methods lacked MPC’s predictive capability essential for DSM pricing environments.
Recent work has focused on MPC tuning. For example, \cite{DENG2024339} applied heuristic search to tune an LSTM-based MPC, while \cite{9482646} used sample average approximation to evaluate performance under uncertainty. \rv{However, these studies primarily target specific problems such as model structure improvement or building design parameter selection under uncertainties, rather than the direct tuning of cost- and comfort-related hyperparameters within the MPC controller itself. Moreover, the associated optimization procedures can exhibit sensitivity to initialization and may converge to suboptimal solutions.}

Bayesian optimization (BO) alleviates this issue and is effective for black-box objectives such as total energy cost~\citep{lu2021mpc}. Yet, many BO-based MPC methods either cannot directly support black-box constraints~\citep{lu2021mpc} or rely on manually identified reference models~\citep{LU2021107491}, limiting scalability.
This gap is critical in DSM-based building control, where minimizing electricity cost under complex pricing must also ensure long-term comfort. \rv{Due to model inaccuracies, limited horizons and nonlinear comfort metrics, these objectives can only be assessed reliably by the closed-loop performance level over extended periods. As a result, controller tuning becomes a black-box optimization problem, for which conventional MPC design and tuning methods are ill suited.} While constrained BO frameworks such as CONFIG~\citep{xu2023constrained} can handle such objectives and constraints, their use in MPC tuning for DSM-based building control remains largely unexplored. 
This raises a key question: can these methods reliably deliver cost-efficient and comfort-compliant control across diverse real-world electricity contracts? 
\rv{While prior studies have investigated MPC tuning for specific modeling or design objectives, the application of constrained Bayesian optimization to tune cost- and comfort-related MPC hyperparameters under DSM pricing has not yet been systematically studied.}

To address this, we propose a performance-oriented MPC design framework based on the CONFIG algorithm, a constrained Bayesian optimization approach that ensures global optimality while satisfying black-box constraints. The framework automates MPC hyperparameter tuning for black-box objectives and constraints. Validation shows that the optimized, computationally efficient MPC achieves performance comparable to a computationally intensive MPC. A case study across 12 real Belgian electricity contracts further demonstrates substantial cost savings and offers practical insights for contract selection while reliably maintaining occupant comfort.

\begin{figure*}[ht]
     \centering
     \includegraphics[width = 0.8\textwidth]{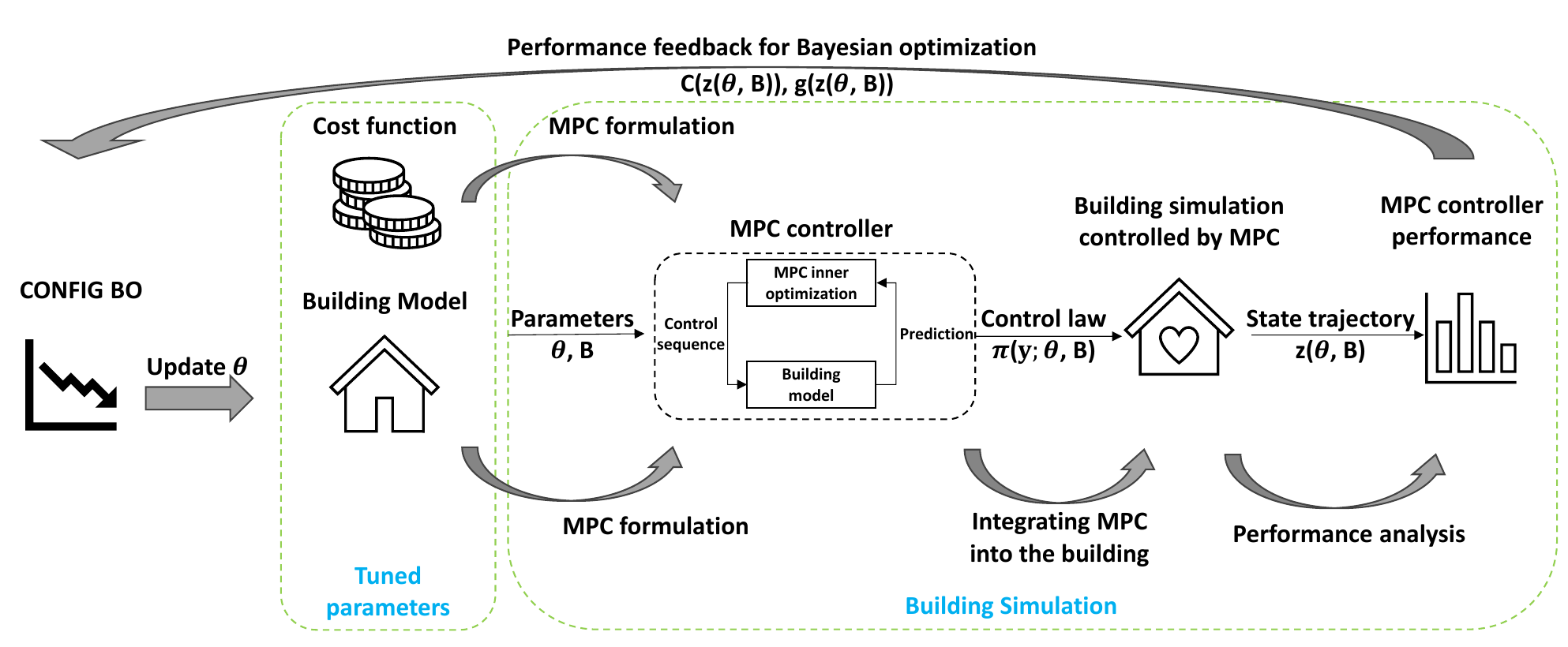}
     \caption{An overview of the approach for developing the performance-oriented MPC controller tuning in building control
     }
     \label{fig: pipeline}
\end{figure*}

\section{Problem statement}
\label{sec: prob}
This paper presents a hyperparameter offline tuning approach to optimize the performance of an MPC building controller under various electricity billing contracts. Let $\theta$ be a vector of tuning parameters for an MPC controller and $\mathcal{B}=  \{B_1, B_2, \allowbreak \ldots, B_n\}$ be the set of potential electricity contracts. The total electricity cost over a considered time period T is denoted as $C^{\star}$ and the contract selection problem to be solved is
\begin{align}
    C_{\min} &= \min_{B \in \mathcal{B}} C^{\star}(B)
    \label{eq: main_prob}
\end{align}

We define the control law as $\pi(y; B, \theta)$, which is a function of the state $y$ and is parameterized by the energy contract $B$ and the controller parameters $\theta$. Given the control law $\pi$, we define the closed-loop state and input trajectories as $\mathbf{z}(B, \theta) = \{(y_0,u_0),\dots,\allowbreak (y_T,u_T)\}$ over the tuning period $T$. The details of the control law implementation as a generic MPC controller are given in section.~\ref{sec:MPC_design}.

We then define the \emph{controller tuning problem} as
\begin{align}
    C^\star(B) = \min_\theta&\  C(\mathbf{z}(\theta, B), B)\\
    \text{s.t.}&\ g(\mathbf{z}(\theta, B)) \le 0
\end{align}
where $C(\mathbf{z}, B)$ is the total financial cost of the trajectory $\mathbf{z}$ under the contract $B$, and $g$ measures the comfort of the occupants of the building.

\section{MPC Tuning in Building Control}
\subsection{Overview}
Our performance-oriented methodology uses historical data from an arbitrary local controller to identify a parametric building model, focusing solely on dynamics rather than usage patterns. As shown in Fig.~\ref{fig: pipeline}, we formulate an MPC controller with predefined objectives and constraints for offline tuning, with online deployment reserved for future work. Using building simulation data, we evaluate MPC performance metrics like electricity costs and thermal comfort, which the CONFIG algorithm uses to optimize MPC parameters in a continuous cycle. This approach leverages digital twin technology~\citep{Grieves2017DigitalTM} to simulate and optimize building operations without disrupting the physical system.
\rv{In practice, constructing a highly accurate, building-specific digital twin can be challenging and costly, particularly for general residential customers. Rather than requiring a customized model for each deployment, the proposed framework allows the digital twin used for offline tuning to be adapted from existing models of similar building topologies and scales. Key characteristics such as geographic location, weather profiles, occupancy schedules, and approximate building size can be adjusted to reflect the target building. In practice, the digital twin used for offline tuning does not need to be an exact replica of the target building. It mainly needs to preserve the dominant input–output behavior relevant to long-horizon cost and comfort evaluation (e.g., weather/occupancy disturbances and actuator characteristics). Therefore, a digital twin adapted from an existing model of a similar building typology can provide a sufficiently representative offline evaluation environment for tuning while avoiding the burden of detailed, building-specific model development~\citep{SHEN2024114478}.}

\rv{In practice, the workflow proceeds as follows:
\begin{enumerate}
\item Historical operational data are collected from an existing local controller operating in the building.
\item The collected data are used to identify a building model that captures the dominant thermal dynamics. This model may be purely data-driven or hybrid, integrating basic physical relationships to improve accuracy and reduce dependence on large datasets.
\item An MPC formulation with predefined objectives and constraints is constructed based on the identified building model.
\item Sensitive hyperparameters are selected (e.g., via sensitivity analysis), and offline optimization is performed using the digital twin to evaluate long-horizon performance metrics.
\item The optimized hyperparameters are deployed in the real building controller.
\item Optionally, the procedure can be repeated periodically as part of recommissioning to adapt to system changes or evolving operating conditions.
\end{enumerate}
}

With this framework, the optimal hyperparameter set and the corresponding electricity cost for a given billing contract can be obtained efficiently. Finally, by comparing the optimal results of various contracts, we provide guidance for consumers to select the most economical option and offer tailored control recommendations.

\subsection{CONFIG for MPC Controller Tuning}

Controller parameter optimization presents a black-box problem due to unclear relationships between hyperparameters, objectives, and constraints. We address this using Bayesian Optimization with the CONFIG algorithm, which efficiently handles complex constraints and provides high-probability optimality guarantees~\citep{xu2023constrained}. \rv{CONFIG enforces constraints by optimizing lower confidence bound surrogates of both the objective and the constraints, rather than relying on hard safety sets or tuned penalty terms. In contrast to penalty-based~\citep{LU2022895} or primal–dual approaches~\citep{XU2024103212} commonly used in MPC auto-tuning, CONFIG does not require tuning penalty coefficients or dual step sizes and naturally supports infeasibility detection. This formulation permits controlled constraint violations during learning while providing explicit bounds on cumulative constraint violations and convergence to the constrained optimum.} This approach enables optimal parameter selection that minimizes costs while satisfying operational requirements under specific billing contracts.

In our CONFIG Bayesian Optimization framework, we model the unknown objective function $C(z(\theta, B), B)$ and constraint $g(z(\theta, B))$ using Gaussian process surrogates. Specifically, we employ $\mathcal{GP}(0, k_i(\cdot, \cdot))$ for $i \in \{0\}\cup\{1\}$ and an \rv{independent and identically distributed} Gaussian noise model with variance $\lambda$.
The objective measurement $h_{0, k} = C(z(\theta_k, B), B)$ and constraint measurement $h_{1, k} = g(z(\theta_k, B))$ with sample $\theta_k$ at iteration $k$ are both corrupted by sub-Gaussian noise $\sigma$ respectively. Let $\Theta_k$ denote the sample sequence $(\theta_1, \theta_2, \dots, \theta_k)$. For the Gaussian process, the following functions for $\theta$ and $\theta'$ are introduced,
\begin{subequations}
\small
    \begin{align}
        \mu_{0,k}(\theta) &= k_0(\theta_{1:k},\theta)^{\top}(K_{0,k}+\lambda I)^{-1}h_{0,1:k} \\
         k_{0,k}(\theta, \theta')&=k_0(\theta, \theta') -k_0(\theta_{1:k}, \theta)^{\top}(K_{0,k}+\lambda I)^{-1}k_0(\theta_{1:k}, \theta')\\
        \sigma_{0,k}^2(\theta)&=k_{0,k}(\theta, \theta)
    \end{align}
\end{subequations}
where $k_0(\theta_{1:k},\theta)=[k_0(\theta_1, \theta), k_0(\theta_2, \theta), \dots,k_0(\theta_k, \theta) ]^{\top}$, $K_{0,k}=(k_0(\theta, \theta')_{\theta,\theta' \in \Theta_k})$ and $h_{0,1:k}=[h_{0,1}, h_{0,2}, \dots, h_{0,k}]^{\top}$. Through the same approach, $\mu_{1,k}(\cdot), k_{1,k}(\cdot)$ and $\sigma_{1,k}(\cdot)$ can be obtained. 

With the Gaussian process posterior distribution functions, the performance oriented optimization is applied as illustrated in Algorithm.~\ref{alg:lcb2}. A weighting parameter $\beta^{1/2}_{i,k}$ is introduced to balance the decision of whether to explore further unknown area or to exploit at the current spot. $K$ represents the total iterations of the CONFIG algorithm and $\Theta$ denotes the space of hyperparameters to be explored and evaluated. Rather than directly solving the problem with unknown objective and constraint functions, the method solves an auxiliary problem that substitutes these functions with their lower confidence bound surrogates.
A theoretical guideline for selecting the value of $\beta_{i,k}^{1/2}$ to guarantee the global optimality is introduced and proved in \cite{xu2023constrained}.

\begin{algorithm}[htbp!]
\small
	\caption{CONFIG for Controller Tuning}
	\begin{algorithmic}[1]
    \For{$k\in[K]$}
      \If
     { $\min_{\theta\in\Theta}(\mu_{1,k}(\theta)-\beta^{1/2}_{1,k}\sigma_{1,k}(\theta))>0$}
     \State
     \textbf{Declare infeasibility}. ~\label{alg_line:declare_inf} 
     \EndIf
     \State\label{alg_line:aux_prob} 
     Update controller parameters with 
     \[
     \begin{aligned}
     \theta_k\in{\arg}&\min_{\theta\in \Theta}\quad\mu_{0,k}(\theta)-\beta^{1/2}_{0,k}\sigma_{0,k}(\theta)\label{eq:lcb}\\
     \textbf{ subject to }& \mu_{1,k}(\theta)-\beta^{1/2}_{1,k}\sigma_{1,k}(\theta)\leq0,.
     \end{aligned}
     \]
     
     \State Run building simulation to get measurements of $C(z(\theta_k, B), B)$ and $g(z(\theta_k, B))$.
     \State Update Gaussian process posterior mean and covariance with the new evaluations added. 
    \EndFor
	\end{algorithmic}
\label{alg:lcb2}
\end{algorithm}

\section{Case study settings}
\subsection{Simulation platform}
\label{sec:simu_plat}
The simulation platform for this work is the Building Optimization Performance Tests (BOPTEST~\cite{boptest}), specifically using the \textit{bestest\_hydronic\_heat\_pump} test case. This represents a Brussels-based residential building modeled as a single thermal zone with BESTEST case 900 parameters. The case study considers a single-zone residential building in Brussels, Belgium, measuring $12 \times 16 \times 2.7$ m. \rv{Internal walls are configured such that there are around 12 rooms in the building. The building further contains $24$ $m^2$ of windows on the south facade. The underlying building geometry contains multiple rooms. However, the BOPTEST interface provides a single aggregated zone temperature, and we therefore adopt a single-zone control abstraction.} The dwelling is occupied by five residents before 7 AM and after 8 PM on weekdays, and full-time on weekends. Space heating is provided by a 15 kW air-to-water modulating heat pump connected to a floor heating system, where the heating power is controlled through a modulation signal ranging from 0 (off) to 1 (full capacity). 

\rv{Heat pumps are increasingly deployed in European residential buildings~\citep{EHPA2020HeatPumpsSound}. In this context, the selected test case provides a representative residential heat-pump-based heating setup, while deliberately adopting a simplified single-zone abstraction. Despite this simplification, the model captures the key thermal dynamics, occupancy effects, and actuator behavior relevant for control-oriented evaluation. Importantly, the proposed optimization framework is model-agnostic and is not tied to this specific representation. It can be directly extended to higher-fidelity or multi-zone building models given an appropriate simulation environment or digital twin.}

\subsection{Tuning objective and constraints for CONFIG}
\label{sec: CONFIG-obj-cons}
\vspace{-0.5em}

\textit{\textbf{Objective Function}} This work targets monthly electricity bill reduction for Belgian households, a more complex challenge than energy minimization due to dynamic, nonlinear pricing structures. To demonstrate our performance-oriented optimization framework's efficacy, we specifically formulate electricity cost minimization as the primary control objective.

To provide a realistic assessment, we examined actual electricity policies in Vlaanderen, Belgium, selected for their inclusion of capacity tariffs (charges based on monthly peak power consumption measured at 15-minute intervals). This creates a nonlinear pricing structure that poses challenges for controller integration. For analysis, we categorized charges into three components: time-of-use energy charges, capacity tariffs, and fixed fees. A total of 12 distinct contracts were evaluated, with detailed specifications provided in Table.~\ref{tab:elec_contrats}.

\vspace{-1.5em}
\begin{align}
    h_{0, k} &= C(z(\theta_k, B), B) \nonumber \\
    & = (\sum^{N-1}_{t=0} u_{agg, t}p^B_{elec,t}) +  p^B_{capa}*u_{peak} +p^B_{fixed}
    \label{eq: objective}
\end{align}
\vspace{-1em}

\rv{For a given electricity contract $B$, we introduce $u_{peak}$ to be the max peak electricity power used in 15 minutes of a month, $p^B_{elec}$ to be the normal electricity price, $p^B_{capa}$ to be the capacity tariff, $p^B_{fixed}$ to be the fixed charge, $u_{agg,t}$ to be the measured electricity power used in step $t$, aggregated across all actuators in the building, $N$ to be the total simulation time step and then the objective of a month for CONFIG algorithm can be defined as Eq.~\eqref{eq: objective}. Note that $u_{agg,t}$ in the electricity cost computation refers to the aggregated real power consumption across all actuators in the building and $u_{peak}$ is the maximum of this aggregated power. However in the present case study, we consider a single-zone model with only one heat pump and therefore no aggregation is required and the expressions reduce accordingly.}

\textit{\textbf{Constraint Function}} Optimal building control requires addressing constrained optimization problems. In addition to direct temperature constraints, this work incorporates a Predicted Mean Vote (PMV) constraint to evaluate thermal comfort, using \cite{Fanger1970-ye}'s widely-adopted seven-point scale from -3 (cold) to +3 (hot) with zero representing thermal neutrality. 
\rv{While PMV provides a comprehensive and standardized measure of thermal comfort, it is a nonlinear function of multiple environmental and occupant-related variables and is therefore not straightforward to embed directly within a linear MPC formulation. Instead, PMV can be easily evaluated externally and enforced as an explicit constraint within the proposed performance-oriented optimization framework. Temperature bounds are used as soft constraints to define an admissible operating envelope inside MPC, while thermal comfort is ultimately ensured through the PMV constraint in our proposed framework.}

This project implements a relaxed PMV constraint where, following standard guidance in \cite{noauthor_2007-xq}, the PMV should remain between -0.5 and 0.5. \rv{While the standards specify instantaneous comfort ranges, limited time-aggregated violations are also discussed in the comfort literature~\citep{sourbron2011evaluation}.} To allow for practical violations, we use a cumulative distribution function (CDF) of the PMV values recorded during the simulation. The constraint requires that the 80\% of the absolute PMV values during occupied hours (denoted as $PMV_{CDF}^B$ for contract $B$) is less than 0.5, ensuring comfort standards are met at least 80\% of the time, as formalized in Eq.~\eqref{eq: cons}. \rv{The 80\% compliance criterion adopted in this work does not correspond to a specific certification class and should not be interpreted as strict compliance with a particular standard. Since comfort targets vary across applications and standards define multiple comfort categories, we treat the compliance threshold as a configurable parameter that can be adapted to user preferences or stricter compliance requirements. Importantly, we report the realized comfort metrics for all controllers to ensure transparency, and the method aims to reduce cost subject to the chosen comfort requirement rather than by implicitly relaxing comfort.} \rv{Note that this CDF-based PMV constraint is again difficult to incorporate directly into a standard MPC formulation, especially when the MPC prediction horizon does not span the full evaluation period required by occupants or billing contracts.This further highlights the advantage of the proposed framework, which enables comfort constraints defined over long horizons to be enforced through performance-oriented optimization rather than embedded directly within the MPC.}

\vspace{-1em}
\begin{equation}
    h_{1, k} = g(z(\theta_k, B)) = PMV_{CDF}^B - 0.5
    \label{eq: cons}
\end{equation}
\vspace{-1em}

In this research, only one constraint is implemented. However, both the CONFIG algorithm and the PMV integrated CDF constraint can handle multiple constraints, such as adding a constraint requiring that the absolute PMV value at 50\% is smaller than 0.2.

To calculate the PMV, an open-source Python tool package called pythermalcomfort~\citep{Tartarini2020-ap} is employed. For the PMV calculation, we use a metabolic rate of $1.2$ (for schooling), relative humidity of $50\%$, air velocity of $0.1 m/s$, and a clothing index of $1.0$ (for winter). \rv{To be noted, for PMV computation, we assume the room is thermally well mixed, i.e. air velocity is slow, radiant temperature asymmetry is small and surface temperatures are close to air temperature, such that operative temperature can be approximated by air temperature~\citep{ASHRAE552021}. This approximation is further supported by large-scale empirical studies showing that the difference between air temperature and operative or mean radiant temperature is typically small and of limited practical relevance for control purposes under standard indoor conditions~\citep{DAWE2020109582}. This commonly adopted approximation enables a consistent and reproducible PMV evaluation within the ARX-based workflow. While it may affect the absolute PMV value under strong radiant asymmetry, the optimization framework and the controller comparisons in this study use the same PMV computation consistently.}

 \begin{table*}[!t]
\centering
\caption{12 Different electricity contract prices in Belgium, Vlaanderen}
\begin{threeparttable}
\begin{tabular}[t]{ccccccc}
\toprule
\multirow{2}{*}{\textbf{Name}} & \multirow{2}{*}{\textbf{Single }}\tnote{1}& \multicolumn{2}{c}{\textbf{Day and night}\tnote{2}} & \multirow{2}{*}{\textbf{Capacity tariff}}\tnote{3} & \multirow{2}{*}{\textbf{Fixed charge}} & \multirow{2}{*}{\textbf{Pricing case}}\tnote{4}\\
&&Day&Night&\\
\midrule
Easy Dynamic\tnote{5} & -&0.307&0.270&3.35&11.2&ddd\\
(with capacity tariff)&0.297&-&-&3.35&11.2&dds\\
\midrule
Easy Dynamic & -&0.329&0.292&-&19.6&dnd\\
(no capacity tariff)&0.319&-&-&-&19.6&dns\\
\midrule
Easy Static Low & -&0.278&0.253&3.35&11.2&sdd\\
(with capacity tariff)&0.270&-&-&3.35&11.2&sds\\
\midrule
Easy Static Low & -&0.300&0.275&-&19.6&snd\\
(no capacity tariff)&0.392&-&-&-&19.6&sns\\
\midrule
Easy Static High\tnote{6} & -&0.307&0.270&3.35&11.2&sdd\\
(with capacity tariff)&0.297&-&-&3.35&11.2&sds\\
\midrule
Easy Static High & -&0.329&0.292&-&19.6&snd1\\
(no capacity tariff)&0.319&-&-&-&19.6&sns1\\

\bottomrule
\end{tabular}
\begin{tablenotes}   
        \footnotesize  
        \item[1] Unit: For Single and Day and Night, the unit is \euro/kWh. For Capacity tariff, the unit is \euro/kW/month. For Fixed charge, the unit is \euro/month. The total monthly electricity bill is: (Single or Day and Night)$\times$ time of use power + Capacity tariff $\times$ peak power per month + Fixed charge.
        \item[2] Day time: 7:00 to 22:00, Monday to Friday; Night time is from 22:00 to 7:00, Monday to Friday, and both weekends
        \item[3] A charge based on the peak power used every 15 minutes anytime during the month.
        \item[4] Composed of three letters: First:`d'-dynamic, `s'-static; Second: `d'-with capacity tariff, `n'-no capacity tariff; Third:`d'-day and night, `s'-single tariff. 
        \item[5] Dynamic means price for each month is different. The prices shown in the table are for November of 2023. Please find more details from different contracts in \cite{engie} .
        \item[6] Static price is decided by the month to sign the contract. Therefore, we have chosen 2 static contracts to compare. '1' denotes the statics contract with higher price.

      \end{tablenotes}  
\end{threeparttable}
\label{tab:elec_contrats}
\end{table*}%

\subsection{MPC Design}
\label{sec:MPC_design}

\rv{We adopt a linear data-driven ARX model for the MPC implementation as a simple and well-established baseline. Such models are widely used in the building sector for MPC due to their effectiveness and efficiency~\citep{DRGONA2020190,khosravi2024,shi2024}. At the same time, the proposed optimization and tuning framework is model-agnostic and does not rely on any ARX-specific properties.
When extending to multi-zone buildings, stronger inter-zone coupling can make MIMO identification and maintenance more challenging, motivating structured/sparse model designs (e.g., constraining coupling using zone adjacency)~\citep{JinSparseARX}. Alternatively, low-order gray-box or physics-based predictors~\citep{joe2023model,liu2025building} can be employed for large-scale deployments without modifying the proposed workflow.
We also note that a fixed ARX model may require sufficiently rich datasets to capture different operating regimes. A practical remedy is to update the predictor online using the latest persistently exciting data, as supported by recent adaptive data-driven building-control studies~\citep{shi2024disturbance,shi2024}.}

\textbf{\textit{Basic Configuration}} \rv{Before introducing the MPC formulation, we emphasize that thermal comfort is enforced at the outer level in the CONFIG via the PMV-based compliance constraint (Eq.~\eqref{eq: cons}). The temperature bounds introduced in this section are internal MPC constraints used to maintain feasibility and operational robustness which are not the primary comfort metric.} We formulate an economic-like MPC controller to minimize electricity costs, where the objective is based on the price times the modulated signal, rather than real power consumption. At time step $t$, let $p_t$ be the electricity price, $y_t$ the indoor temperature, and $u_t$ the modulated heat pump control input. The building dynamics follow Eq.~\eqref{eq: pre} using ARX coefficients $A(q)$ and $B(q)$, with input vector $\mathbf{u}_{ARX}$. The $\mathbf{u}_{ARX}$ is comprised of heat pump signal, outdoor temperature, and solar radiation, subject to constant delay $t_d$.
Constraints define feasible sets $\mathcal{Y}$ for temperature $y_t$ (Eq.~\eqref{eq: cons_y}) and $\mathcal{U}$ for control input $u_t$ (Eq.~\eqref{eq: cons_u}). The heat pump input is bounded as $u_t \in [0, u_{\text{max}}]$ to limit peak power. We set different temperature constraints for occupied and unoccupied scenarios. In this way, $[T^{occupy}_{lb}, T^{occupy}_{ub}]$ and $[T^{unoccupy}_{lb}, T^{unoccupy}_{ub}]$ are two constraints for indoor temperature in occupied hours and unoccupied hours respectively as mentioned in Section.~\ref{sec:simu_plat}. A slack variable $\epsilon$ with penalty coefficient $S$ enforces soft constraints, while $R$ penalizes control effort.
The quadratic programming formulation in Eq.~\eqref{eq: mpc_form}, where $\mathbb{B}$ is a unit ball in an appropriate norm, shows MPC performance as a black-box function of parameters including $\mathcal{U}$, $R$, $A(q)$ and etc.

\vspace{-1em}
\begin{subequations}
\begin{align}
\centering
   \textbf{QP: }\; \objective^*_{mpc}(y,u)&=\mathop{\min}_{y,u}\sum^{N-1}_{t=0} Rp_tu_t+\epsilon_t'S\epsilon_t  \label{eq: cost_func} \\    
\text{subject to:}&\nonumber\\
&  A(q)y(t)=B(q)u_{ARX}(t-t_d) \label{eq: pre}\\
& y_t \in \mathcal{Y} \oplus \epsilon \mathbb{B} \label{eq: cons_y}\\
& \mathcal{Y} = [T^{occupy}_{lb}, T^{occupy}_{ub}]  \text{ if occupied} \label{eq:QP_Y}\\
& \mathcal{Y} = [T^{unoccupy}_{lb}, T^{unoccupy}_{ub}]  \text{ otherwise}\nonumber \\
& u_t \in \mathcal{U} \label{eq: cons_u} \\
& \mathcal{U} = [0, u_{max}] \label{eq: cons_U}
\end{align}
\label{eq: mpc_form}
\end{subequations}
\vspace{-1em}

\textbf{\textit{Minimum Operating Power Issue}} After integrating the basic QP MPC into the building simulation, a minimum operating power (MOP) issue was identified as shown in Fig.~\ref{fig:mask_mpc}. While zero input (modulated heat pump power) draws zero power, even slight non-zero inputs activate components like heat pump fans, causing power spikes to around 1 kW. Unlike bang-bang control which avoids intermediate states, MPC can produce small non-zero inputs leading to frequent low-power activations. This results in inefficient operation at minimum power without delivering significant thermal output.
Since the periodic PRBS signal used for ARX model identification cannot capture this MOP behavior, the issue must be resolved at the control level.

\begin{figure}[!t]
    \centering
    \begin{tikzpicture}
    \begin{axis}[
        width=0.85\columnwidth,height=6cm,  
        xlabel={Modulated heat pump power},
        ylabel={Electricity power (W)},
        legend style={font=\footnotesize},
        title={Modulated input and actual electricity power in January},
        xmin=-0.1, xmax=1.1,  
        ymin=-100, ymax=3700,  
        grid=major, grid style={dashed,gray!30},  
        xtick={0, 0.3, 0.8, 1.0},  
        ytick={0, 1000, 1700, 2700, 3500},  
        legend pos=north west,  
        ]

        \addplot[
        only marks, mark=diamond, mark size=3pt, blue] 
        table[meta = elec]
        {fig/baseline_scatter_data.dat};
        \addlegendentry{Bang-bang control};

        \addplot[
        only marks, mark=x, mark size=1pt, orange] 
        table[meta = elec]
        {fig/mask_scatter_data.dat};
        \addlegendentry{QP MPC control};

        \addplot[domain=0:0.3, purple, line width=3pt] {0};
        \addplot[line width=3pt, purple] coordinates {(0.3,0) (0.3,1700)};
        \addplot[domain=0.3:0.8, purple, line width=3pt] {1700 + 2000 * (x - 0.3)};
        \addplot[domain=0.8:1, purple, line width=3pt] {2700};
        \addplot[line width=3pt, purple] coordinates {(1,0) (1,2700)};
        \addlegendentry{Simplified MPC}

    \end{axis}
\end{tikzpicture}
    \caption{MOP illustration and simplified-MPC masking concept
    }
    \label{fig:mask_mpc}
\end{figure}

\begin{table*}[ht]
\begin{center}
\caption{Approximation methods to include MOP issue in the MPC}
\begin{tabular}{ cl } 
\hline
\textbf{Approximation Case} & \textbf{Details} \\
\hline
\multirow{2}{*}{MIQP MPC} & $\mathrm{U}=\{(u, P_{elec}) \quad | \quad 0 \leq u\leq u_{max}*\alpha, \alpha \in \{0, 1\}, P_{elec}= \phi \cdot u + \gamma \alpha \}$\\ 
&  $u_{mpc} = mask(u) = u$\\
\hline
\multirow{2}{*}{Simplified MPC} & $\mathrm{U}=\{(u, P_{elec}) \quad | \quad 0 \leq u\leq u_{max}, P_{elec}=  u   \}$\\ 
& $u_{mpc} = mask(u) = 0 \quad if \quad u < u_{low} \quad else \quad u$ \\
\hline
\end{tabular}
\label{tab: MIQP_and_Mask}
\end{center}
\end{table*}

\textbf{\textit{MPC Design}} We set up a MPC targeted to solve the MOP issue as shown in Eq.~\ref{eq: mpc_MOP}. Instead of directly input $u_t$ to the simulator, we have introduced an extra variable called $u_{mpc, t}$ for time step $t$. The mapping from $u_t$ to $u_{mpc, t}$ is defined by function $mask()$. Then we introduce two approximation methods to include MOP issue in the MPC as shown in Table.~\ref{tab: MIQP_and_Mask}.

\begin{subequations}
\small
\begin{align}
\centering
   \textbf{MOP:}\; \objective^*_{mpc}(y,u)&=\mathop{\min}_{y,u,\alpha}\sum^{N-1}_{t=0} Rp_tP_{elec,t}+\epsilon_t'S\epsilon_t  \label{eq: cost_MIQP} \\    
\text{subject to:} \nonumber\\
&  \eqref{eq: pre},\eqref{eq: cons_y},\eqref{eq:QP_Y} \\
& u_t, P_{elec, t} \in \mathrm{U}\\
& u_{mpc, t} = mask(u_t)
\end{align}
\label{eq: mpc_MOP}
\end{subequations}

\vspace{-1em}

For the MIQP MPC case, the relationship between the modulated signal, $u$, and the real electricity power, $P_{elec}$, can be approximated as a mixed-integer constraint, i.e. $P_{elec}= \phi \cdot u + \gamma$, where $\gamma$ denotes the system minimum operating power and $\phi$ represents the power to increase per unit of $u$ increase. This formulation introduces integer variables, making the control problem a Mixed-Integer Quadratic Programming (MIQP) problem. A binary variable $\alpha$ (which can be either 0 or 1) is used to introduce switching behaviors or on-off decisions. These mechanisms allow the optimization to incorporate both continuous control inputs and discrete decision-making, enhancing the flexibility and applicability of the control strategy. To be noted, as the identification of the ARX model remains the same, $u_{ARX}$ is still composed of the modulated heat pump input $u_{t}$, and the other environmental variables.
For convenience, the MOP problem formulation using the MIQP MPC approximation is referred to as MIQP MPC in the following paper.

Given the computational expense of MIQP problems and the challenges of solving them with open-source solvers like SCIP, particularly on low-cost commercial hardware, we propose a Simplified MPC approach to enable widespread adoption in residential settings. This method applies a mask to the standard MPC to circumvent the minimum operating power issue, as illustrated in Fig.~\ref{fig:mask_mpc} and Table.~\ref{tab: MIQP_and_Mask}. The masked value $u_{mpc, t}$ is then used for the ARX model modulated heat pump input in the next iteration. The mask's lower bound eliminates small input values that trigger inefficient minimum power operation, while its upper bound constrains peak power to reduce demand charges in certain DSM programs and overall energy use. Both bounds are tuned via our optimization framework and maintain the problem as a tractable quadratic program, making it suitable for typical households. This formulation is referred to as Simplified MPC hereafter.

\rv{
As discussed in section.~\ref{sec: CONFIG-obj-cons}, the realized monthly electricity cost under a given contract is evaluated according to Eq.~\eqref{eq: objective}, which includes the normal energy tariff, the capacity tariff, and the fixed charge.
In the Simplified MPC formulation, only a modulated-signal-based energy cost term is included explicitly in the objective, consistent with the QP MPC as in Eq.~\eqref{eq: mpc_form}.  With the lower bound $u_{low} > 0$, the product of electricity price and the modulated control signal provides a sufficiently accurate approximation of the normal real electricity cost, even in the presence of the minimum operating power issue. The capacity-tariff-related cost is instead addressed implicitly through the upper bound $u_{max}$: tighter limits on $u_{max}$ constrain instantaneous power usage and are likely to reduce the maximum 15-minute aggregated power over the billing period, i.e., $u_{peak}$ in Eq.~\eqref{eq: objective}. The fixed charge is not included in the MPC objective, as it is contract-dependent and cannot be influenced by control decisions. 
 }

\textbf{\textit{Complexity Comparison}} In this study, the MPC problem was formulated using the CVXPY package and solved with both MOSEK and SCIP solvers. While the commercial MOSEK solver offers superior performance, the open-source SCIP solver was employed to demonstrate the computational advantages of the Simplified MPC over the MIQP formulation. Under the same pricing scheme, we compared the average simulation time per control step during January, as summarized in Table.~\ref{tab:solve time}. All simulations used a 36-step horizon, a 10th-order linear data-driven ARX model with a scalar output and three inputs, over a one-month simulation period with 15-minute control resolution.

Most simulations were performed on a computing cluster with 128 GB RAM. To demonstrate the impracticality of MIQP MPC for commercial building control, we also tested on a personal computer with 16 GB RAM, which has less memory than the computing cluster but can still be better than a low-cost chip. Results show that MIQP solving is substantially more time-consuming than QP for both MOSEK and SCIP solvers. Notably, SCIP solving MIQP on a PC often becomes unresponsive, exhausting available RAM for extended periods at certain steps. While solving time limits can be imposed to maintain operation, this leads to suboptimal solutions. Given these computational constraints, Simplified MPC emerges as clearly preferable to MIQP MPC when both deliver comparable control performance.
\begin{table}[ht]
\small
\begin{center}
\caption{Average time for solving one step on three runs for January}
\begin{tabular}{ ccc } 
\hline
\multirow{2}{*}{\textbf{Case}} & \multicolumn{2}{c}{\textbf{Average solving time [sec]}}\\
& \textbf{Cluster} & \textbf{PC} \\
\hline
MOSEK QP& 0.015& 0.011 \\ 
MOSEK MIQP& 0.13& 0.089\\ 
SCIP QP& 0.15& 0.233\\ 
SCIP MIQP& 0.80& $ \infty $ \\
\hline
\end{tabular}
\label{tab:solve time}
\end{center}
\end{table}

\vspace{-1em}
The comparison highlights the infeasibility of deploying a complex MIQP MPC controller in households due to the need for an expensive commercial solver. In contrast, the Simplified MPC can be solved with a free solver, though it is slower. However, since building management system inputs are typically updated every 15 minutes or longer, there is sufficient time for a free solver to solve the QP MPC problem.

\subsection{Tuning Variables in MPC}
\label{sec: tuning_variables}
 
Four hyperparameters were selected for tuning with the Simplified MPC: $T^{occupy}_{lb}$, $T^{unoccupy}_{lb}$, $u_{max}$, and $u_{low}$. The lower bounds, $T^{occupy}_{lb}$ and $T^{unoccupy}_{lb}$, are particularly significant, as the indoor temperature trajectory tends to align with the lower bounds without a tracking reference in the objective, thus heavily influencing individual comfort. $u_{max}$ is crucial for limiting the electricity power applied, thus contributing to cost reduction. $u_{low}$ addresses the system minimum operating power problem, as discussed in the previous section. The tuning set is presented in Table.~\ref{tab:tuning}, and optimal values are selected from the specified ranges.

\begin{table}[!h]
\begin{center}
\caption{Tuning variables for Simplified MPC}
\begin{tabular}{ ccccc } 
\hline
Variable & $u_{low}$ & $u_{max}$ & $T^{occupy}_{lb}$ & $T^{unoccupy}_{lb}$ \\
\hline
Range & [0, 0.8] & [0.8, 1] & [21, 23] & [15, 18] \\
\hline
\end{tabular}
\label{tab:tuning}
\end{center}
\end{table}

\begin{figure*}[!ht]
\centering
        \begin{tikzpicture}
\begin{axis}[
        width=0.9\textwidth, height=5cm,
        ylabel={Electricity cost (\euro)},
        ymin=0, ymax=400,
        title ={Electricity costs for different months with different controllers (Pricing case: ddd)},
        symbolic x coords={November, December, January, February},
        xtick=data,
	legend style={at={(0.5,-0.2)},
	anchor=north,legend columns=-1},
	ybar = 4,
        bar width = 20pt,
        nodes near coords,
        enlarge x limits=0.15,
        every node near coord/.append style={font=\scriptsize},
        ymajorgrids=true, 
]
\addplot coordinates {(November, 275.54) (December, 343.86) (January,345.49) (February, 266.61)};
\addlegendentry{Baseline}
\addplot coordinates {(November, 244.01) (December, 297.60) (January,304.25) (February, 237.91)};
\addlegendentry{ QP MPC}
\addplot coordinates {(November,205.18) (December, 256.56) (January,273.54) (February, 209.38)};
\addlegendentry{MIQP MPC}

\addplot+[
  error bars/.cd,
  y dir=both,
  y explicit,
  error bar style={draw=red, line width=0.8pt}
]
coordinates {
  (November,201.41) +- (0,0.79)
  (December,255.60) +- (0,0.47)
  (January,272.26)  +- (0,4.19)
  (February,207.42) +- (0,1.23)
};
\addlegendentry{Tuned MPC}

\node[font=\scriptsize, text=red, align=center, xshift=36pt]
  at (axis cs:November,160) {+ -\\0.79};

\node[font=\scriptsize, text=red, align=center, xshift=36pt]
  at (axis cs:December,160) {+ -\\0.47};

\node[font=\scriptsize, text=red, align=center, xshift=36pt]
  at (axis cs:January,160) {+ -\\4.19};

\node[font=\scriptsize, text=red, align=center, xshift=36pt]
  at (axis cs:February,160) {+ -\\1.23};

\end{axis}
\end{tikzpicture}
        \caption{Electricity cost comparison of three different controllers. The proposed Tuned MPC can achieve greater performance after tuning in our proposed
hyperparameter optimizing framework than the MIQP MPC controller. \rv{Red values inside the Tuned MPC bars are the standard deviations over 4 runs.}}
        \label{fig:mpc_month}
\end{figure*}

\begin{figure*}[!ht]
\centering
        \begin{tikzpicture}

\begin{axis}[
    ylabel={Temperature ($^{\circ}C$)},
    ymin=14, ymax=28,
    width=0.9\textwidth, height=4cm,
    legend style={font=\footnotesize,fill=white, fill opacity=0.8, draw=none},
    xmin=386.5, xmax=395,
    xtick={387, 388, 389, 390, 391, 392, 393, 394},
    xticklabels={Monday, Tuesday, Wednesday, Thursday, Friday, Saturday, Sunday},
    legend pos = south east,
]
\addplot[blue] table [x=time, y=Baseline, col sep=space, mark =none]{fig/Jan_temp_data.dat};
\addlegendentry{Indoor temperature (Baseline)}
\addplot[orange] table [x=time, y=Mask, col sep=space, mark =none]{fig/Jan_temp_data.dat};
\addlegendentry{Indoor temperature (Tuned MPC)}
\addplot[green] table [x=time, y=QP, col sep=space, mark =none]{fig/Jan_temp_data.dat};
\addlegendentry{Indoor temperature (QP MPC)}
\addplot[purple] table [x=time, y=MIQP, col sep=space, mark =none]{fig/Jan_temp_data.dat};
\addlegendentry{Indoor temperature (MIQP MPC)}
\addplot[red, dash dot] table [x=time, y=low, col sep=space, mark =none]{fig/Jan_temp_data.dat};
\addlegendentry{Indoor temperature (Constraint)}
\addplot[red, dash dot] table [x=time, y=high, col sep=space, mark =none]{fig/Jan_temp_data.dat};
\end{axis}
\end{tikzpicture}
        \caption{Performance comparison of four different controllers. One week segment performance of different controllers in January (One month total cost , Baseline:345 \euro, QP MPC:304.25 \euro, MIQP MPC:273.54 \euro, Tuned MPC:272.26 \euro). \rv{The dashed temperature-constraint lines show the offline-optimized bounds used by the Tuned MPC and are therefore constant over the plotted period; the resulting tuned parameters are: $(u_{low}, u_{max}, T^{occupy}_{lb},T^{unoccupy}_{lb}) = (0.8,0.959,21.163^{\circ}C, 15^{\circ}C)$}}
        \label{fig:mpc_Jan_sec}
\end{figure*}

\section{Offline optimization with BOPTEST}

\subsection{Baseline controllers}
\rv{This case study employs three baseline controllers for comparison. The first is a rule-based controller provided by the BOPTEST platform, which operates with fixed temperature setpoints of $21.2^\circ C$ during occupied hours and $20.5^\circ C$ during unoccupied hours. The second is an MIQP-based MPC (Table.~\ref{tab: MIQP_and_Mask}) that explicitly addresses the minimum operating power of the heat pump using empirically selected, manually tuned parameters. These parameters were obtained through repeated trial-and-error experiments aimed at maintaining acceptable comfort levels while reducing energy costs, a process that requires repeated closed-loop simulations when performed offline (as in this study), and, if one were to tune online, could initially cause comfort violations before a feasible configuration is identified. The MIQP MPC employs modulated heat pump power bounds of [0,1] and temperature setpoints of $21.5^\circ C$ for occupied hours and $16^\circ C$ for unoccupied hours. These bounds were obtained via the above trial-and-error tuning and are reported here as the final selected values. The third is a QP-based MPC formulation (Eq.~\eqref{eq: mpc_form}) that neglects minimum operating power constraints but uses the same temperature setpoints and power bounds as the MIQP MPC. Consequently, the performance difference observed between the QP MPC and the tuned Simplified MPC can be attributed largely to the hyperparameter tuning process rather than differences in temperature settings or actuation limits.}

\subsection{Experiments design}
\vspace{-1em}
To enhance MPC controller performance, simulations were run for one-month periods in November through February, capturing significant weather variations for the ARX model. Results for Tuned MPC are averaged across four initial conditions:
$(u_{low}, u_{max}, T^{occupy}_{lb},T^{unoccupy}_{lb}) \in \{(0.2, 1, 22, 17),(0.5, 1, 22, 16),(0.2, 0.8, 23 ,17),(0.5, 0.8,\newline 21, 18)\}$. For CONFIG, the objective variance is set to $4$, and the constraint variance to $0.04$. Kernel length scales, determined by the tuning variable range in Table~\ref{tab:tuning}, are $[0.4, 0.1, 1.5, 1.5]$. We use a constant $\beta^{1/2}_{1, k}=5$ for simplicity, and grid search is employed to solve the CONFIG problem in its low-dimensional space, with $50$ total iterations ($K$).

Two experiments were designed to address the objectives. First, we compared the performance of the Tuned MPC against three baseline controllers, optimizing four hyperparameters (as detailed in Section~\ref{sec: tuning_variables}). In the second experiment, we tuned the same hyperparameters across 12 pricing cases, requiring 12 optimization runs to identify the best performance for each case. Finally, we provided guidance for local households of similar scale.

\begin{table*}[!h]
\centering
\caption{Optimized electricity bills under different pricing methods for different months}
\begin{threeparttable}

\begin{tabular}[t]{ccccccccc}
\toprule
\multirow{3}{*}{\textbf{Pricing case}} & \multicolumn{8}{c}{\textbf{Electricity bills (\euro/month)}}\\
&  \multicolumn{2}{c}{\textbf{November}}&  \multicolumn{2}{c}{\textbf{December}}& \multicolumn{2}{c}{\textbf{January}} & \multicolumn{2}{c}{\textbf{February}} \\
& Tuned MPC& QP MPC & Tuned MPC& QP MPC& Tuned MPC& QP MPC& Tuned MPC& QP MPC\\
\midrule
ddd &201.41&244.01 &255.60&297.60 &272.26&304.25 &\textbf{207.42}&237.91 \\
dds &211.96&259.16 &269.05&314.89 &284.96&322.41 &217.58& 251.00 \\
dnd &212.13&256.28 &270.84&316.13 &288.37&324.73 &224.70&254.03 \\
dns &223.57&271.83 &284.22&333.76 &302.88& 342.90 &230.47&267.11 \\
sdd &\textbf{190.22}&228.97 &\textbf{248.95}&290.12 &\textbf{272.14}&328.13 &222.62& 252.92 \\
sds &195.08&237.42 &257.19&301.15 &282.33&348.81 &227.22&262.68 \\
snd &201.82&242.00 &266.43&308.82 &289.97&348.59 &235.59&269.01 \\
sns &205.11&251.12 &272.28&319.95 &299.85&369.31 &239.77&278.86 \\
sdd1 &201.31&243.78 &267.59&310.39 &291.28&328.13 &236.75&270.37 \\
sds1 &212.38&259.22 &280.53&329.10 &308.17&348.59  &246.53&286.75 \\
snd1 &213.54&256.22 &281.56&329.24 &310.13&348.81 &246.65&286.35 \\
sns1 &\textbf{222.72}&271.89 &\textbf{296.23}&347.82 &\textbf{325.84}&369.31 &\textbf{259.86}& 302.81 \\
\hline
SP\tnote{1}& 33.35&- & 47.28&- & 53.70&- & 52.44&- \\
SR\tnote{2}& 14.94\%&- & 15.96\%&- & 16.48\%&- & 20.18\%&- \\
\bottomrule
\end{tabular}
\begin{tablenotes}   
        \footnotesize  
        \item[1] SP: Saving potential, worst case bill - best case bill, unit: \euro/month.
        \item[2] SR: Saving ratio, (worst case bill - best case bill)/worst case bill, unit:1. 
\end{tablenotes}  
\end{threeparttable}
\label{tab: pricing_results}
\end{table*}%

\subsection{Results and discussions}

\subsubsection{Details of optimization}
\label{sec:basic_results}

\vspace{-1em}

Firstly, a comparative analysis of the final electricity costs was conducted for the baseline controller, the QP MPC, the computationally intensive MIQP MPC controller, and the tuned Simplified MPC (Tuned MPC) under the same pricing case (ddd as shown in Table.~\ref{tab:elec_contrats}), as illustrated in Fig.~\ref{fig:mpc_month}. It is evident that in comparison to the baseline controller, all three MPC controllers significantly reduce the final electricity costs for different months. \rv{The reported Tuned MPC costs are mean ± std over four tuning runs with different initial conditions (each run reporting its own best-found parameters within the iteration budget). The relative standard deviation (std/mean) of the Tuned MPC cost remains below 2\% across all four months, indicating high reproducibility and low sensitivity to initialization, with consistently similar best-found performance across runs.}

Fig.~\ref{fig:mpc_Jan_sec} highlights that the MPC controller's cost savings stem from its superior prediction capabilities, including its enhanced knowledge of temperature constraints and electricity prices compared to the baseline rule-based controller. While the cost savings between the QP MPC and the Tuned MPC is the profit achieved by hyperparameter tuning. In fact, the hyperparameter tuning contributes to as much cost reduction as the predictive ability of MPC which illustrates the vitality to perform hyperparameter tuning on MPC controllers in building control and the superior effects of our optimization framework. 
The Tuned MPC, in particular, has the potential to reduce these costs by up to 26.90\% compared with baseline controller and up to 17.46\% compared to the QP MPC. For all months, the costs of the Tuned MPC are consistently smaller than those of the MIQP MPC. It can be demonstrated that the proposed Tuned MPC can achieve greater performance after tuning in our proposed hyperparameter optimizing framework than the MIQP MPC controller. However, the computational cost of the Tuned MPC is much lower than that of the MIQP MPC, as demonstrated in Table.~\ref{tab:solve time}.

\begin{figure}[t]
    \centering
\begin{tikzpicture}
    \begin{axis}[
        width=0.85\columnwidth,height=5cm,  
        xlabel={$|PMV|$},
        ylabel={Percentage},
        legend style={font=\footnotesize},
        title={CDF of $|PMV|$ during occupied hours},
        xmin=-0.1, xmax=1.1,  
        ymin=-0.1, ymax=1.1,  
        xtick={0, 0.2, 0.4, 0.6, 0.8, 1.0},  
        ytick={0, 0.2, 0.4, 0.6, 0.8, 1.0},  
        yticklabels={0\%, 20\%, 30\%, 60\%, 80\%, 100\%},
        legend pos=south east,  
        ]
        \addplot[blue] table [ x=x, y=y, col sep=space, mark =none] {fig/base_cdf_data.dat};
        \addlegendentry{Baseline};
        \addplot[green] table [x=x, y=y, col sep=space, mark =none] {fig/MIQP_cdf_data.dat};
        \addlegendentry{MIQP MPC};
        \addplot[orange] table [x=x, y=y, col sep=space, mark =none] {fig/Mask_cdf_data.dat};
        \addlegendentry{Tuned MPC};

        \addplot[red, dash dot, thick] coordinates {(0.5,-0.1) (0.5, 1.1)};
        \addplot[red, dash dot, thick] coordinates {(-0.1,0.8) (1.1, 0.8)};
        \addlegendentry{Constraint};
    \end{axis}
\end{tikzpicture}
    \caption{Cumulative distributive function of absolute PMV during occupied hours in January. Constraint: 80\% of the absolute PMV values during occupied hours are smaller than 0.5.  
    }
    \label{fig:mpc_cdf}
\end{figure}

To clearly illustrate how the proposed optimization framework enforces comfort constraints, we have compared the comfort index CDF of the three controllers as shown in Fig.~\ref{fig:mpc_cdf}. From the CDF of absolute PMV values during the occupied hours in January, it can be observed that while all three controllers satisfy the comfort constraint (80\% of the absolute PMV values during occupied hours are smaller than 0.5), the Tuned MPC is most effective in achieving this result through the precise tuning of temperature constraints and the mask in the Simplified MPC. Together with Fig.~\ref{fig:mpc_month} and Fig.~\ref{fig:mpc_Jan_sec}, it can be concluded that the Tuned MPC applies more accurate temperature constraints after optimization, resulting in further cost savings while maintaining comfort constraints. 
\rv{Importantly, the proposed framework does not achieve cost reductions by relaxing comfort requirements. Instead, comfort is explicitly enforced through the PMV-based constraint, and the optimized controllers operate strictly within this limit.}
Overall, with significantly lower commissioning costs in terms of solver and hardware requirements compared to the MIQP MPC, the Tuned MPC achieves comparable or even superior performance.

\subsubsection{Comparison of electricity contracts}

After validating the performance-oriented MPC tuning method, we analyzed 12 electricity contracts (Table.~\ref{tab:elec_contrats}) to provide guidance for Belgian consumers seeking cost-optimal contracts. Results in Table.~\ref{tab: pricing_results} demonstrate that contract selection significantly impacts costs, with optimal choices saving up to €53.70 monthly (20.18\% reduction compared to worst-case contracts). The performance gap between Tuned MPC and QP MPC further underscores the critical importance of hyperparameter optimization in building control efficiency. All Tuned MPC results reflect contract-specific optimizations, confirming our method's effectiveness in minimizing electricity expenses.

From the Tuned MPC results, we can also indicate that it would be beneficial if the customers can choose a different contract for each month, however, it is impractical as there has been a minimum of one year duration for all contracts. The analysis shows that tariff structures combining day and night pricing with capacity charges consistently yield the best performance across both dynamic and fixed-rate schemes. Among these, the dynamic ddd contract and the fixed sdd contract stand out: ddd performs optimally in February, while sdd is most effective in other months. While sdd appears generally advantageous, its benefit relies on securing sufficiently low fixed rates at the time of contract signing. If electricity prices are high, the final cost of electricity will be less favorable than that of the dynamic option, similar to the sdd1. Considering electricity price uncertainty, the dynamic ddd contract is recommended, as it provides stable performance without exposure to unfavorable fixed-rate conditions.

While manually tuned QP MPC may suggest similar contracts, it can be suboptimal in complex scenarios, such as with electric vehicle charging, where high peak demand increases capacity charges. In such cases, QP MPC might favor contracts with high capacity tariffs but low time-of-use rates, ignoring peak demand costs. In contrast, Tuned MPC optimizes for total cost, including capacity charges, and provides more appropriate recommendations in these situations.

\section{Conclusion}

This paper tackles two key challenges in building control: selecting cost-optimal electricity contracts and improving the performance of conventional MPC designs. We propose a performance-oriented optimization strategy based on the CONFIG algorithm for constrained black-box optimization. Using the BOPTEST digital twin platform, the method conducts offline optimization to identify MPC hyperparameters that minimize costs while satisfying comfort constraints.

A Belgian residential case study shows that a properly tuned simple MPC can match the performance of computationally intensive alternatives, reducing monthly electricity bills by up to 26.9\% and by 17.46\% compared to an manually tuned MPC during heating months. The framework also optimizes contract selection, with the best option achieving 20.18\% savings relative to the worst case.

While offline optimization effectively addresses practical implementation challenges, future work should extend it toward online adaptation, enabling dynamic performance estimation and continuous improvement in real time. \rv{Owing to the properties of the CONFIG algorithm, the risk and magnitude of constraint violations could be reduced during online operation by adopting a conservative confidence parameter $\beta$~\citep{pmlr-v37-sui15} and restricting the search to a neighborhood around the offline-optimized solution. Besides, the online extension could also be combined with a supervisory safety mechanism (e.g., fallback controller~\cite{shi2024disturbance}, constraint tightening, safety filter~\cite{wabersich2018safe}) to prevent unacceptable comfort violations. This enables a conservative online refinement phase that preserves comfort and safety while allowing gradual performance improvement.}

\begin{ack}
\vspace{-0.5em}
This research was supported by the Swiss National Science Foundation under NCCR Automation, grant agreement 51NF40\_180545.
\end{ack}

\section*{DECLARATION OF GENERATIVE AI AND AI-ASSISTED TECHNOLOGIES IN THE WRITING PROCESS}
\vspace{-0.5em}
During the preparation of this work the author(s) used ChatGPT in order to improve language clarity and conciseness. After using this tool, the author(s) reviewed and edited the content as needed and take(s) full responsibility for the content of the publication.

\bibliography{ifacconf}

\end{document}